\documentclass[journal=jpcbfk,manuscript=article]{achemso}
\SectionNumbersOn 

\usepackage[T1]{fontenc}

\usepackage{graphicx}
\usepackage{float}

\usepackage{mathtools}
\usepackage{amssymb}
\usepackage{derivative}
\usepackage{dsfont}

\usepackage{siunitx}
\usepackage{xcolor}
\usepackage[hidelinks]{hyperref}
\usepackage{seqsplit}
\usepackage[title]{appendix}

\DeclareMathOperator{\E}{\mathds{E}}
\DeclareMathOperator{\MSE}{MSE}
\DeclareMathOperator{\MSR}{MSR}

\newcommand{\CaRMSD}[1]{\operatorname{RMSD}_\text{#1}}
\newcommand{\dt}{\epsilon}
\newcommand{\1}{\mathds{1}}
\newcommand{\bmin}{\wedge}
\newcommand{\bmax}{\vee}
\newcommand{\comp}{\mathsf{c}}

\newcommand{\peq}{\mathrel{\phantom{=}}}

\title{An exact multiple-time-step variational formulation for the committor and the transition rate}
\author{Chatipat Lorpaiboon}
\affiliation{Department of Chemistry and James Franck Institute, University of Chicago, Chicago, Illinois 60637, United States}
\author{Jonathan Weare}
\affiliation{Courant Institute of Mathematical Sciences, New York University, New York, New York 10012, United States}
\author{Aaron R. Dinner}
\affiliation{Department of Chemistry and James Franck Institute, University of Chicago, Chicago, Illinois 60637, United States}
\email{dinner@uchicago.edu}

\begin{document}
\maketitle

\begin{abstract}
For a transition between two stable states, the committor is the probability that the dynamics leads to one stable state before the other. It can be estimated from trajectory data by minimizing an expression for the transition rate that depends on a lag time.  We show that an existing such expression is minimized by the exact committor only when the lag time is a single time step, resulting in a biased estimate in practical applications.  We introduce an alternative expression that is minimized by the exact committor at any lag time. 
 The key idea is that, when trajectories enter the stable states, the times that they enter (stopping times) must be used for estimating the committor and transition rate instead of the lag time.
Numerical tests on benchmark systems demonstrate that our committor and transition rate estimates are much less sensitive to the choice of lag time.  
We show how further accuracy for the transition rate can be achieved by combining results from two lag times.
We also relate the transition rate expression to a variational approach for kinetic statistics based on the mean-squared residual and discuss further numerical considerations with the aid of a decomposition of the error into dynamic modes. 
\end{abstract}

\section{Introduction}

For a reaction in which a system transitions from a reactant state $A$ to a product state $B$, the probability that the dynamics leads to $B$ before $A$ from a microscopic state $x$---known as the commitment probability or splitting probability, or simply the committor $q(x)$---is a natural (arguably optimal \cite{du_transition_1998,hummer_transition_2004,e_transition_2005,e_towards_2006,e_transition-path_2010,berezhkovskii_diffusion_2013,banushkina_optimal_2016}) reaction coordinate.
 Transition states can be identified directly from the dynamics as the states with $q(x)=0.5$ (i.e., equal probability of going to $A$ and $B$), and integration along the committor yields transition rates and related quantities \cite{e_transition_2005,e_towards_2006,e_transition-path_2010,strahan_long-time-scale_2021}.  Collective variables that correlate with the committor yield informative projections of the free energy \cite{ma_automatic_2005,hu_two-step_2008}, and they can be used to guide enhanced sampling methods efficiently; in fact, the committor can serve as an optimal controller for sampling the transition path ensemble \cite{singh_splitting_2024}.  

Directly computing the committor for a microscopic state $x$ involves simulating many trajectories from $x$ to the $A$ and $B$ states. Though machine learning can be used to make efficient use of this information \cite{ma_automatic_2005,peters_obtaining_2006,peters_extensions_2007,hu_two-step_2008,jung_machine-guided_2023}, directly computing the committor is prohibitively computationally expensive for many systems.
To avoid this expense, many algorithms have been proposed which exploit the Markov structure of the underlying process to learn an estimate of the committor using a dataset of short trajectories (most neither beginning in $A$ nor ending in $B$) initialized from states distributed throughout the state space\cite{thiede_galerkin_2019,strahan_long-time-scale_2021,strahan_predicting_2023,strahan_inexact_2023,li_semigroup_2022,chen_discovering_2023,aristoff_fast_2024,evans_computing_2022,evans_computing_2023,song_finite_2025,mitchell_committor_2024,lucente_coupling_2022,lucente_committor_2022,jacques-dumas_data-driven_2023,rotskoff_active_2022,kang_computing_2024}.

Specifically, when the dynamics satisfy detailed balance and the trajectory data are from an equilibrium simulation, the committor can be approximated by minimizing the loss function
\cite{chen_discovering_2023}
\begin{equation}\label{eq:reactiveflux}
  \Phi = \frac{1}{2\tau}\E[(q(X_\tau)-q(X_0))^2],
\end{equation}
 subject to boundary conditions $q(x) = 0$ for $x \in A$ and $q(x) = 1$ for $x \in B$.
Here, $X_t$ is the state of the Markov process at time $t$, with $X_0$ drawn from the equilibrium distribution. The parameter $\tau$ is known as the lag time.
While \eqref{eq:reactiveflux} can be viewed as a finite-time approximation of the loss function $\E[\lVert \nabla q \rVert^2]$, which has also been minimized  subject to the same boundary conditions as well as continuity constraints to approximate the committor function \cite{khoo_solving_2019,li_computing_2019,li_semigroup_2022,rotskoff_active_2022,chen_committor_2023}, \eqref{eq:reactiveflux} can also be derived by minimizing the steady-state flux between $A$ and $B$ \cite{roux_string_2021,roux_transition_2022,he_committor-consistent_2022}.
Below, we obtain this expression from transition path theory \cite{e_transition_2005,e_towards_2006,e_transition-path_2010}.

Minimization of \eqref{eq:reactiveflux} yields the exact committor function when the lag time is a single time step.
However, often analyses are performed for a projection onto a subset of variables (e.g., the positions without the velocities, the solute but not solvent degrees of freedom, distances between selected atoms, etc.), and  the dynamics of these variables are generally non-Markovian at such short lag times \cite{zwanzig_nonequilibrium_2001}.
At longer lag times, the Markov assumption better holds for projected dynamics \cite{bowman_introduction_2013}, but then, as we discuss below, the exact committor no longer minimizes \eqref{eq:reactiveflux}  because \eqref{eq:reactiveflux} does not take into account that trajectories can enter and exit $A$ and $B$.
In this article, we address this issue by introducing a variational expression that yields the exact committor and transition rates for arbitrary lag times.

The article is organized as follows.
We review committors in Section \ref{sec:committor} and outline the derivation of a multiple-time-step estimator for the transition rate in Section \ref{sec:main_rate}; details of the derivation are provided in Appendix \ref{sec:tpt_rate}.
Based on the estimator for the transition rate, we introduce a multiple-time-step variational expression in Section \ref{sec:loss} and discuss its relation to \eqref{eq:reactiveflux} and the loss function in Ref.~\citenum{li_semigroup_2022}.  Then we discuss how the error in the committor can be quantified and introduce an improved estimator using two lag times in Section \ref{sec:metrics:mse}, with further discussion in Appendices \ref{sec:modedecomp} to \ref{sec:msr}.  We describe the neural network architectures that we use in Section \ref{sec:nns} and the systems and data that we use for our numerical tests in Section \ref{sec:systems}. We show how our formulation improves committor and transition rate estimates in Sections \ref{sec:committor_results} to \ref{sec:mse_results}.  We discuss possible directions for future research in Section \ref{sec:discussion} and additional numerical considerations in Appendix \ref{sec:overfitting}.
 An implementation of this method, and its application to the examples in this work, is available at \url{https://github.com/dinner-group/evcn}.

\section{Methods}

Our main theoretical contributions are as follows:
\begin{itemize}
    \item
    Without assuming detailed balance, we derive \eqref{eq:tpt_rate}, an exact multiple time step expression for the transition rate.  Importantly,
    \eqref{eq:tpt_rate} involves the times of last exit from and first entry to $A$ and $B$ when they differ from the lag times (i.e., we ``stop'' trajectories at the boundaries).
    \item
    Assuming detailed balance, we use \eqref{eq:tpt_rate} to obtain \eqref{eq:evcn_loss}, an exact variational expression for the transition rate that can be minimized for the committor.  In addition to using the stopping times as described above, we show how the stopping considerations naturally give rise to terms in the loss that penalize violation of the boundary conditions.
    \item
    We introduce \eqref{eq:approx_rate}, a more accurate estimator for the transition rate, and \eqref{eq:approx_mse}, an estimator for the mean squared error in the committor; these estimators are linear combinations of \eqref{eq:evcn_loss} at two different lag times.
\end{itemize}

Our analysis is based on a discrete-time stationary ergodic Markov process $X_t$ with time step $\dt$.
Continuous-time dynamics can be obtained by taking the limit $\dt \to 0$.
We assume that states $A$ and $B$ are disjoint, and define the transition region $D = (A \cup B)^\comp$.

\subsection{Committor}\label{sec:committor}

The forward committor $q(x)$ is the probability that a system at $x$ will enter $B$ before $A$.
The backward committor $\bar{q}(x)$ is the probability that a system at $x$ exited $A$ after $B$.
Mathematically, they are defined as
\begin{align}
  q(x) & = \E[\1_B(X_{S_t}) \mid X_t = x],
  \label{eq:qp_def} \\
  \bar{q}(x) & = \E[\1_A(X_{\bar{S}_t}) \mid X_t = x],
  \label{eq:qm_def}
\end{align}
where
\begin{align}
  \label{eqn:Stdef}
  S_t & = \min \{ s \mid s \ge t , X_s \notin D \}, \\
  \label{eqn:Sbartdef}
  \bar{S}_t & = \max \{ s \mid s \le t, X_s \notin D \},
\end{align}
and
\begin{equation}
  \1_C(x) =
  \begin{cases}
    1 & \text{if } x \in C, \\
    0 & \text{otherwise}.
  \end{cases}
\end{equation}
When the dynamics obey detailed balance, the forward and backward committors are related by $\bar{q} = 1 - q$.

Committors can be estimated from single-step trajectories by solving the boundary value problems:
\begin{align}
  q(x) & = \E[q(X_{t+\dt}) \mid X_t = x],
  \label{eq:qp_bvp_1} \\
  \bar{q}(x) & = \E[\bar{q}(X_{t-\dt}) \mid X_t = x],
  \label{eq:qm_bvp_1}
\end{align}
for $x \in D$, with boundary conditions $q(x) = \1_B(x)$ and $\bar{q}(x) = \1_A(x)$ for $x \notin D$.
These boundary value problems follow from applying the law of total expectations (at times $t-\dt$ and $t+\dt$, respectively) to \eqref{eq:qp_def} and \eqref{eq:qm_def} when $x \in D$.
We emphasize that \eqref{eq:qp_bvp_1} and \eqref{eq:qm_bvp_1} are exact only when $\dt$ is a single time step.

Defining $(t+\tau) \bmin S_t = \min\{t+\tau,S_t\}$ and $(t-\tau) \bmax \bar{S}_t = \max\{t-\tau,\bar{S}_t\}$, for multiple time steps, the boundary value problems are instead
\begin{align}
  q(x) & = \E[q(X_{(t+\tau) \bmin S_t}) \mid X_t = x],
  \label{eq:qp_bvp} \\
  \bar{q}(x) & = \E[\bar{q}(X_{(t-\tau) \bmax \bar{S}_t}) \mid X_t = x],
  \label{eq:qm_bvp}
\end{align}
with the same boundary conditions as the single time step case. 
These boundary value problems follow from applying the law of total expectations (at times $(t+\tau) \bmin S_t$ and $(t-\tau) \bmax \bar{S}_t$, respectively) to \eqref{eq:qp_def} and \eqref{eq:qm_def}.
They are exact for arbitrary $\tau \ge 0$ because $S_t$ and $\bar{S}_t$ stop the process when it hits $A$ or $B$  (see Refs.~\citenum{strahan_long-time-scale_2021} and~\citenum{tuchkov_error_2025} for further discussion).
Unlike the single time step case, \eqref{eq:qp_bvp} and \eqref{eq:qm_bvp} also hold for $x \notin D$.

\subsection{Transition rate}\label{sec:main_rate}

Denoting the number of transition paths from $A$ to $B$ within the trajectory segment $X_{[t,t']}$ by $N_{AB}(X_{[t,t']})$, the transition rate $\Phi$ is the mean of  $N_{AB}$ per unit time:
\begin{equation}
  \Phi = \lim_{T \to \infty} \frac{1}{2 T} N_{AB}(X_{[-T,T]}).
\end{equation}
This rate is symmetric with respect to $A$ and $B$ and differs from the standard transition rate from $A$ to $B$ defined in transition path theory, which accounts for the time the process spends in state $A$ \cite{e_transition_2005,e_towards_2006,e_transition-path_2010}.
In Appendix \ref{sec:tpt_rate}, we use transition path theory to derive an expression for the transition rate in terms of committors and finite-length trajectories:
\begin{equation} \label{eq:tpt_rate}
  \Phi = \frac{1}{\tau} \E[f(X_{[0,\tau]})],
\end{equation}
where
\begin{align}
  f(X_{[t,t']}) = {}
  & \bar{q}(X_t) \1_D(X_{t' \bmin S_t}) q(X_{t'}) (\xi(X_{t'}) - \xi(X_t))
  \nonumber \\
  & + \bar{q}(X_t) \1_B(X_{t' \bmin S_t}) (1 - \xi(X_t))
  \nonumber \\
  & + \1_A(X_{t \bmax \bar{S}_{t'}}) q(X_{t'}) (\xi(X_{t'}) - 0)
  \nonumber \\
  & + N_{AB}(X_{[t,t']}),
  \label{eq:tpt_rate_telescope}
\end{align}
and $\xi$ is a reaction coordinate that satisfies $\xi(x) = 0$ for $x \in A$ and $\xi(x) = 1$ for $x \in B$.
The terms on the right hand side of \eqref{eq:tpt_rate_telescope} correspond respectively to contributions from transition paths that
start before $t$ and end after $t'$;
start before $t$ and end within $[t,t']$;
start within $[t,t']$ and end after $t'$;
and both start and end within $[t,t']$.
 The importance of the stopping times $S_t$ and $\bar{S}_{t'}$ cannot be overemphasized---they are needed to account correctly for trajectories that enter and exit $A$ and $B$ \cite{strahan_long-time-scale_2021}.

In the case that $\tau=\epsilon$, \eqref{eq:tpt_rate_telescope} reduces to
\begin{equation}\label{eq:tpt_rate_1}
  f(X_{[t,t+\dt]}) = \bar{q}(X_t) q(X_{t+\dt}) (\xi(X_{t+\dt}) - \xi(X_t)).
\end{equation}
Expression \eqref{eq:tpt_rate_1} can also be derived from the transition path theory formula for the reactive flux through a separating surface between $A$ and $B$, by integrating over level sets of $\xi$ \cite{strahan_long-time-scale_2021}.
We emphasize that \eqref{eq:tpt_rate_telescope} and \eqref{eq:tpt_rate_1} do not rely on detailed balance.


When the dynamics satisfy detailed balance, such that $\bar{q} = 1 - q$, we can manipulate \eqref{eq:tpt_rate} to obtain an expression for the transition rate that is quadratic in the committor.
To this end, we set $\xi = q$ and define the time-reversed trajectory $\bar{X}_{[t,t']} = (X_{t'},X_{t'-\dt},\dots,X_t)$.
Then, we can write \eqref{eq:tpt_rate} as
\begin{align}
  \Phi
  & = \frac{1}{2 \tau} \E[f(X_{[0,\tau]}) + f(\bar{X}_{[0,\tau]})] \\
  & = \frac{1}{\tau} \E[g(X_{[0,\tau]},q)],
  \label{eq:rate_lag_sym}
\end{align}
where
\begin{align}
  g(X_{[t,t']},u) = \frac{1}{2} (
    & \1_D(X_{t' \bmin S_t}) (u(X_{t'}) - u(X_t))^2
    \nonumber \\
    & + \1_{D^\comp}(X_{t' \bmin S_t}) (\1_B(X_{t' \bmin S_t}) - u(X_t))^2
    \nonumber \\
    & + \1_{D^\comp}(X_{t \bmax \bar{S}_{t'}}) (u(X_{t'}) - \1_B(X_{t \bmax \bar{S}_{t'}}))^2
    \nonumber \\
    & + N_{AB}(X_{[t,t']}) + N_{BA}(X_{[t,t']})
  ).
  \label{eq:rate_lag_sym_one}
\end{align}
There is a direct correspondence between the terms on the right hand side of \eqref{eq:rate_lag_sym_one} and those on the right hand side of \eqref{eq:tpt_rate_telescope} and, in turn, the contributions described below \eqref{eq:tpt_rate_telescope} (i.e., contributions from the four possible combinations of starting and not starting and ending and not ending in the time interval), but \eqref{eq:rate_lag_sym_one} includes contributions from  transition paths both from $A$ to $B$ and from $B$ to $A$.
As for \eqref{eq:tpt_rate_telescope}, the stopping times $S_t$ and $\bar{S}_{t'}$ are needed to account correctly for trajectories that enter and exit $A$ and $B$.

When $\tau=\epsilon$, \eqref{eq:rate_lag_sym} reduces to
\begin{equation}
  \Phi = \frac{1}{2 \dt} \E[(q(X_\dt) - q(X_0))^2].
  \label{eq:rate_sym}
\end{equation}
Expression \eqref{eq:rate_sym} was previously obtained by \citeauthor{roux_string_2021} from the formula for the reactive flux through a separating surface. \cite{roux_string_2021}

\subsection{Variational loss functions}\label{sec:loss}

Expressions \eqref{eq:rate_lag_sym} and \eqref{eq:rate_sym}  can be used to optimize the committor and transition rate. 
Given a model $u(x)$ for the committor,  we can define the loss function
\begin{equation}
  \tilde{L}_\tau(u) = \frac{1}{\tau} \E[g(X_{[0,\tau]},u)].
  \label{eq:evcn_loss}
\end{equation}
%
%
We note that this loss function includes a penalty for the boundary conditions:
\begin{align}
  g(X_{[t,t']},u)
  & = g(X_{[t,t']},\hat{u}) \nonumber \\
  & \peq {} + \frac{1}{2} \1_{D^\comp}(X_t) (u(X_t) - \1_B(X_t))^2 \nonumber \\
  & \peq {} + \frac{1}{2} \1_{D^\comp}(X_{t'}) (u(X_{t'}) - \1_B(X_{t'}))^2,
\end{align}
where we have introduced the function $\hat{u}(x)$ that is identical to $u(x)$ for $x\in D$, but satisfies the boundary condition $\hat{u}(x) = \1_B(x)$ for $x\in D^\comp$ exactly. 
We stress that the terms enforcing the boundary conditions arise directly from \eqref{eq:rate_lag_sym_one}.  One could modulate their contributions by multiplying them by a tunable hyperparameter.
The gradient of \eqref{eq:evcn_loss} with respect to parameters $\theta$ is
\begin{align}
  \nabla_\theta \tilde{L}_\tau(u) = \frac{1}{\tau} \E[
    & \nabla_\theta u(X_0) (u(X_0) - \hat{u}(X_{\tau \bmin S_0})) \nonumber \\
    & + \nabla_\theta u(X_\tau) (u(X_\tau) - \hat{u}(X_{0 \bmax \bar{S}_\tau}))
  ],
  \label{eq:evcn_grad}
\end{align}
which is zero when $u = q$ by \eqref{eq:qp_bvp} and \eqref{eq:qm_bvp}.

When $\tau = \dt$, \eqref{eq:evcn_loss} reduces to
\begin{align}
  L_\tau(u) = \frac{1}{2\tau} \E[
    & (\hat{u}(X_\tau) - \hat{u}(X_0))^2
    \nonumber \\
    & + \1_{D^\comp}(X_0) (u(X_0) - \1_B(X_0))^2
    \nonumber \\
    & + \1_{D^\comp}(X_\tau) (u(X_\tau) - \1_B(X_\tau))^2
  ].
  \label{eq:vcn_loss}
\end{align}
Again, the last two terms in the expectation impose a penalty for violating the boundary conditions; 
multiplying them by a tunable hyperparameter would recapitulate the losses in Refs.~\citenum{li_semigroup_2022} and~\citenum{megias_iterative_2025}.
The gradient of \eqref{eq:vcn_loss} is
\begin{align}
  \nabla_\theta L_\tau(u) =
  \frac{1}{\tau} \E[
    & \nabla_\theta u(X_0) (u(X_0) - (\1_B(X_0) + \1_D(X_0) \hat{u}(X_\tau))) \nonumber \\
    & + \nabla_\theta u(X_\tau) (u(X_\tau) - (\1_B(X_\tau) + \1_D(X_\tau) \hat{u}(X_0)))
  ],
  \label{eq:vcn_grad}
\end{align}
which is zero when $\tau = \dt$ and $u = q$ because of \eqref{eq:qp_bvp_1} and \eqref{eq:qm_bvp_1}.
We write these expressions with $\tau$ rather than $\dt$ because we will later use them with $\tau > \dt$.
However, minimizing \eqref{eq:vcn_loss} with $\tau>\epsilon$ does not yield $q$ in the infinite data limit.


In principle, the penalties for violating the boundary conditions are not needed---in Ref.~\citenum{chen_discovering_2023}, the authors minimized a variational expression that enforced boundary conditions directly on the model output and evaluated the loss only in $D$. 
However, we found that this approach often leads to non-physical (near) discontinuities close to $A$ and $B$ in practice.
We have therefore included these boundary penalties, as in Refs.~\citenum{li_semigroup_2022} and~\citenum{megias_iterative_2025}.
Following Refs.~\citenum{chen_discovering_2023} and~\citenum{megias_iterative_2025}, we term \eqref{eq:vcn_loss} the variational committor-based neural network (VCN) loss function; by extension, we term \eqref{eq:evcn_loss} the exact VCN (EVCN).

We now explicitly show that these loss functions are variational by analyzing how errors in the committor affect transition rate estimates.
The EVCN loss function \eqref{eq:evcn_loss} is minimized when $u = q$ and $\tilde{L}_\tau(q) = \Phi$ for all $\tau$.
By adding a perturbation $\eta$ to $q$, we have
\begin{align}
  \tilde{L}_\tau(q + \eta)
  = \Phi + \frac{1}{2\tau} \E[
    & 2 \hat{\eta}(X_0) (q(X_0) - q(X_{\tau \bmin S_0}))
    \nonumber \\
    & + 2 \hat{\eta}(X_\tau) (q(X_\tau) - q(X_{0 \bmax \bar{S}_\tau}))
    \nonumber \\
    & + \1_D(X_{\tau \bmin S_0}) (\hat{\eta}(X_\tau) - \hat{\eta}(X_0))^2
    \nonumber \\
    & + \1_{D^\comp}(X_{\tau \bmin S_0}) \hat{\eta}(X_0)^2
    \nonumber \\
    & + \1_{D^\comp}(X_{0 \bmax \bar{S}_\tau}) \hat{\eta}(X_\tau)^2
    \nonumber \\
    & + \1_{D^\comp}(X_0) \eta(X_0)^2
    \nonumber \\
    & + \1_{D^\comp}(X_\tau) \eta(X_\tau)^2
  ],
  \label{eq:evcn_perturb}
\end{align}
where $\hat{\eta}(x) = \1_D(x)\eta(x)$.
The first two terms inside the expectation of \eqref{eq:evcn_perturb} are zero because of \eqref{eq:qp_bvp} and \eqref{eq:qm_bvp}, and the remaining terms are quadratic, so $\tilde{L}_\tau(u) \ge \Phi$ and \eqref{eq:evcn_loss} is variational.
Because \eqref{eq:vcn_loss} derives from \eqref{eq:evcn_loss}, it follows immediately that \eqref{eq:vcn_loss} is variational as well when $\tau=\epsilon$. When $\tau > \dt$, \eqref{eq:vcn_loss} is not variational and can underestimate $\Phi$.

The loss function in \eqref{eq:evcn_loss} is also closely related to that of Ref.~\citenum{li_semigroup_2022}.
With infinite data, the two differ by a constant (they omit the $N_{AB}$ and $N_{BA}$ terms) and yield the same gradient.
Their method directly estimates the gradient from finite data using only the first term of \eqref{eq:evcn_grad}, which is a consistent approximation of the true gradient but is not the gradient of any loss function on finite data (this issue does not arise in the limit of infinite data because both terms then have the same value due to detailed balance).
By symmetrizing the loss with time-reversed trajectories, we directly compute the gradient from a loss function on finite data.
Moreover, our loss is more interpretable, as its minimum is the transition rate and must be used when that quantity is of interest.

\subsection{Mean squared error and alternative transition rate expression} \label{sec:metrics:mse}

To assess the accuracy of the committor in our numerical tests, we evaluate the mean squared error (MSE):
\begin{equation}
  \MSE(u) = \int (u(x) - q(x))^2 \pi(x) \odif{x},
\end{equation}
where $\pi$ is the stationary distribution.
This is a natural metric for our method because we can rearrange \eqref{eq:evcn_perturb} and take the limit $\tau \to \infty$ to obtain
\begin{equation}
  \lim_{\tau \to \infty} \tau (\tilde{L}_\tau(q + \eta) - \Phi) = \MSE(q + \eta). \label{eq:mse_asymp}
\end{equation}
This result follows from 
$\lim_{\tau \to \infty} \1_D(X_{\tau \bmin S_0}) = \1_D(X_{S_0}) = 0$,
$\lim_{\tau \to \infty} \1_{D^\comp}(X_{\tau \bmin S_0}) = \1_{D^\comp}(X_{S_0}) = 1$,
$\lim_{\tau \to \infty} \1_{D^\comp}(X_{0 \bmax \bar{S}_\tau}) = \lim_{\tau \to \infty} \1_{D^\comp}(X_{\bar{S}_\tau}) = 1$,
and $\hat{\eta}(x)^2 + \1_{D^\comp}(x) \eta(x)^2 =  \1_{D}(x)\eta(x)^2 + \1_{D^\comp}(x) \eta(x)^2= \eta(x)^2$.
As noted above, the first two terms inside the expectation of \eqref{eq:evcn_perturb} are zero because of \eqref{eq:qp_bvp} and \eqref{eq:qm_bvp}.

Because we typically do not have the true committor at each configuration, we must approximate the MSE.
If the dataset consists of long trajectories in equilibrium, we can estimate the MSE using
\begin{align}
  \MSE(u)
  & = \E[ (u(X_0) - \E[\1_B(X_{S_0}) \mid X_0])^2 ]
  \label{eq:mse_2_traj_def} \\
  & = \E[(u(X_0) - \1_B(X_{S_0})) (u(X_0) - \1_B(X_{\bar{S}_0}))].
  \label{eq:mse_2_traj}
\end{align}
As discussed in Refs.~\citenum{strahan_predicting_2023} and~\citenum{strahan_inexact_2023}, \eqref{eq:mse_2_traj_def} cannot be estimated using $X_{S_0}$ (with $S_0$ as defined in \eqref{eqn:Stdef}) alone because of the double sampling problem (that is, $\E[\E[\1_B(X_{S_0}) \mid X_0]^2] \ne \E[\1_B(X_{S_0})^2]$).  We resolve this issue by using \eqref{eq:mse_2_traj}, which involves independent samples conditioned on $X_0$, 
$X_{S_0}$ and $X_{\bar{S}_0}$ (with $\bar{S}_0$ as defined in \eqref{eqn:Sbartdef}).
The MSE estimated from \eqref{eq:mse_2_traj} can be negative due to sampling error, which introduces a constant offset.
This is because the quantity in the expectation is not a squared difference but rather a product of two different differences.
Nonetheless, lower MSE values (closer to negative infinity) generally indicate better performance, even when they are negative.

If the dataset consists of many short trajectories, the MSE cannot be estimated using \eqref{eq:mse_2_traj} because $S_0$ and $\bar{S}_0$ are not generally available.
Motivated by \eqref{eq:mse_asymp}, we define approximations $\tilde{L}_{\tau_1,\tau_2}(u) \approx \Phi$ and $\MSE_{\tau_1,\tau_2}(u) \approx \MSE(u)$ by solving $\tau_1 (\tilde{L}_{\tau_1}(u) - \tilde{L}_{\tau_1,\tau_2}(u)) = \MSE_{\tau_1,\tau_2}(u)$ and $\tau_2 (\tilde{L}_{\tau_2}(u) - \tilde{L}_{\tau_1,\tau_2}(u)) = \MSE_{\tau_1,\tau_2}(u)$:
\begin{align}
  \tilde{L}_{\tau_1,\tau_2}(u)
  & = \frac{\tau_1 \tilde{L}_{\tau_1}(u) - \tau_2 \tilde{L}_{\tau_2}(u)}{\tau_1 - \tau_2},
  \label{eq:approx_rate} \\
  \MSE_{\tau_1,\tau_2}(u)
  & = \frac{\tilde{L}_{\tau_1}(u) - \tilde{L}_{\tau_2}(u)}{1/\tau_1 - 1/\tau_2}.
  \label{eq:approx_mse}
\end{align}
In Appendix~\ref{sec:Lextrapolation}, we show that $\tilde{L}_{\tau_1,\tau_2}(u)$ is an upper bound for $\Phi$ and $\MSE_{\tau_1,\tau_2}(u)$ is a lower bound for $\MSE(u)$.
We relate $\MSE_{\tau_1,\tau_2}(u)$ to the mean squared residual loss function in Ref.~\citenum{strahan_predicting_2023} in Appendix~\ref{sec:msr}.

\subsection{Neural networks}
\label{sec:nns}

For our numerical tests, we approximate the committor using a multilayer perceptron with two hidden layers of 100 neurons each and SiLU activation functions, followed by a sigmoid function that constrains outputs to $[0,1]$.
The model is trained on either the VCN loss function \eqref{eq:vcn_loss} or the EVCN loss function \eqref{eq:evcn_loss}.

Training the committor to convergence often requires many optimization steps (over $10^4$), especially at short lag times.
In this work, we trained models up to 20000 steps due to computational constraints.
However, training often becomes unstable after many steps: the loss spikes and recovers repeatedly without converging.
To mitigate this behavior, we found it essential to use a large batch size (at least 1000 trajectory segments), the Adam-atan2 optimizer \cite{everett_scaling_2024}, and a small learning rate ($10^{-3}$, which is just below the threshold at which training becomes unstable).
 The large batch size increases the consistency in the number of states near the transition states (where the committor varies most rapidly) from batch to batch and, in turn, the consistency of the direction of the gradient.


To make metrics on the training and validation datasets comparable, we partition the trajectories into two subsets.
Five models are trained on each subset and validated on the other.
Each full trajectory is segmented into length $\tau$ trajectory segments using a sliding window.
At each optimization step, 1000 trajectory segments $X_{[t,t+\tau]}$ are uniformly sampled from the training dataset to compute the loss.
Checkpoints are saved after $\{1,2,5,10,20,50,\dots,20000\}$ optimization steps, and the checkpoint with the best loss on the validation dataset is selected.
Unless otherwise indicated, metrics are averaged over the 10 models.

As we discuss further in Appendix \ref{sec:overfitting}, overfitting is a major issue for both VCN and EVCN loss functions.
We thus tested a variety of regularization techniques.
Dropout degraded performance and slowed training.
Batch normalization \cite{ioffe_batch_2015} and layer normalization \cite{ba_layer_2016} initially sped up training, but the models struggled to capture finer details in the committor.
Standard weight decay \cite{loshchilov_decoupled_2019} and weight decay toward initial parameters \cite{kumar_maintaining_2025} both worsened performance.
As such, we do not employ these methods.

\section{Results}

In this section, we describe the neural network architecture and training procedure, introduce the test systems, and then discuss numerical results for the committor and transition rate.  Because long trajectories are available for all systems, we can compare the results directly to the empirical rate, which we compute from the data as
\begin{equation}
    \Phi_\text{emp} = \frac{1}{2 T}[ N_{AB}(X_{[0,T]})+ N_{BA}(X_{[0,T]})],
\end{equation}
where here $T$ is the trajectory length.
Similarly, we compute the empirical committor as
\begin{equation}
    q_\text{emp}(X_t) = \frac{1}{2}[\1_B(X_{S_t}) + \1_B(X_{\bar{S}_t})].
\end{equation}

\subsection{Systems} \label{sec:systems}

We test our method on three molecular systems of increasing complexity: AIB\textsubscript{9}, Trp-cage, and villin.
In this section, we briefly describe these systems and define collective variables (CVs) and states that we use in our analysis.

\subsubsection{\texorpdfstring{AIB\textsubscript{9}}{AIB9}}

AIB\textsubscript{9} is a 9-residue peptide of 2-aminoisobutyric acid (AIB), an achiral, unnatural amino acid that forms both left-handed and right-handed 3\textsubscript{10} helices with equal probability.
We examine the left-to-right helix transition.
The dataset consists of 20 trajectories of \qty{15} {\micro\second} each, which we generated for Ref.~\citenum{strahan_inexact_2023} and analyzed in that study and Ref.~\citenum{lorpaiboon_accurate_2024}.
 We use 10 trajectories for training and the remaining 10 for validation.

\begin{figure*}
  \includegraphics[scale=0.9]{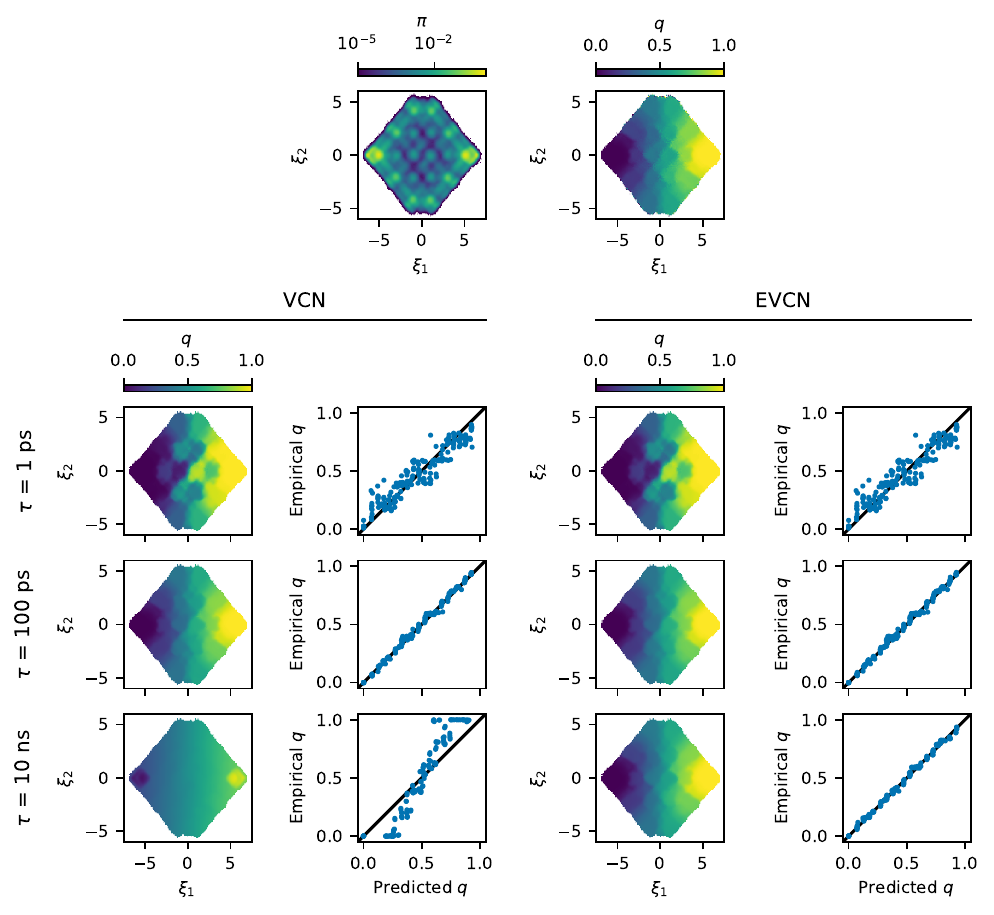}
  \caption{
    CV projections and reliability diagrams for AIB\textsubscript{9}.
    (top left) CV projection of the stationary distribution.
    (top right) CV projection of the empirical committor.
    (left columns of VCN and EVCN) CV projections of the predicted committors.
    (right columns of VCN and EVCN) Reliability diagrams for the predicted committors.
    All results shown are obtained with the dihedral neural-network inputs.
  }
  \label{fig:aib9}
\end{figure*}

For each residue $i$, we define $\gamma_i = -0.8 (\sin(\phi_i) + \sin(\psi_i))$.
The left-handed and right-handed residue conformations have $\gamma_i \approx -1$ and $\gamma_i \approx 1$, respectively, though other configurations may also have these values.
We define the left-handed helix state $A$ and right-handed helix state $B$ as having $(\phi,\psi)$ angles of residues 3--7 within \qty{25}{\degree} of $(\qty{41}{\degree},\qty{47}{\degree})$ and $(\qty{-41}{\degree},\qty{-47}{\degree})$, respectively.
With these state definitions, the left-to-right helix transition has an empirical rate of $9.1 \times 10^{-3}$ \unit{\per\nano\second}.
In Fig.~\ref{fig:aib9}, we show the stationary distribution ($\pi$) and empirical committor projected on the collective variables (CVs) introduced in
Ref.~\citenum{lorpaiboon_accurate_2024}:
\begin{align}
  \xi_1 & = \gamma_3 + \gamma_4 + \gamma_5 + \gamma_6 + \gamma_7, \\
  \xi_2 & = \gamma_3 + \gamma_4 - \gamma_6 - \gamma_7.
\end{align}
The leftmost and rightmost states correspond to states $A$ and $B$, respectively.

We use as inputs to the neural networks two sets of features: the CVs $\xi_1$ and $\xi_2$ (``CVs''),
and the sines and cosines of the $(\phi,\psi)$ dihedral angles for residues 3--7 (``Dihedrals'').

\subsubsection{Trp-cage}

Trp-cage is a designed 20-residue fast-folding protein that has been extensively studied both experimentally and computationally (see Ref.~\citenum{strahan_long-time-scale_2021} and references therein).
Its folded structure consists of an $\alpha$-helix (residues 2--9), a $3_{10}$-helix (residues 11--14), and a polyproline II helix (residues 17--19), which form a cage around Trp6.
We analyze a \qty{208}{\micro\second} trajectory of the K8A mutant (sequence: \seqsplit{DAYAQWLADGGPSSGRPPPS}) at \qty{290}{\kelvin}, saved every \qty{0.2}{\nano\second} \cite{lindorff-larsen_how_2011}.
To generate training and validation data, the trajectory is split in half and treated as two independent segments.

\begin{figure*}
  \includegraphics[scale=0.9]{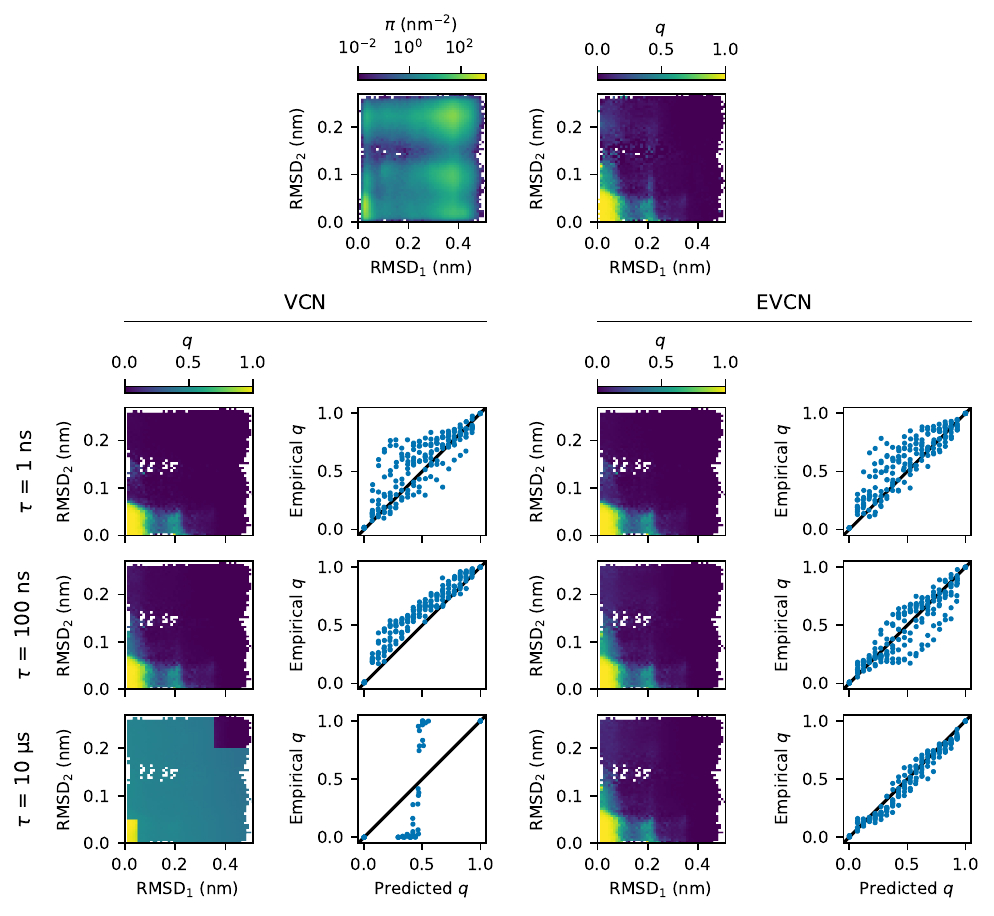}
  \caption{
    CV projections and reliability diagrams for Trp-cage.
    (top left) CV projection of the stationary distribution.
    (top right) CV projection of the empirical committor.
    (left columns of VCN and EVCN) CV projections of the predicted committors.
    (right columns of VCN and EVCN) Reliability diagrams for the predicted committors.
    All results shown are obtained with the dihedral neural-network inputs.
  }
  \label{fig:trpcage}
\end{figure*}

We define three CVs: the root-mean-squared deviation (RMSD) of the C$_\alpha$ atoms to the experimental structure (PDB 2JOF \cite{barua_trp-cage_2008}) of the $\alpha$-helix ($\CaRMSD{1}$), the $3_{10}$-helix ($\CaRMSD{2}$), and the $\alpha$- and $3_{10}$-helices ($\CaRMSD{1,2}$).
Following Ref.~\citenum{strahan_long-time-scale_2021}, we project the results onto $\CaRMSD{1}$ and $\CaRMSD{2}$.
We define the unfolded state $A$ as configurations with $\CaRMSD{1} \ge \qty{0.35}{\nano\meter}$, $\CaRMSD{2} \ge \qty{0.2}{\nano\meter}$, and $\CaRMSD{1,2} \ge \qty{0.4}{\nano\meter}$, and the folded state $B$ as configurations with $\CaRMSD{1} \le \qty{0.05}{\nano\meter}$, $\CaRMSD{2} \le \qty{0.05}{\nano\meter}$, and $\CaRMSD{1,2} \le \qty{0.15}{\nano\meter}$.
With these state definitions, the empirical rate is $7.7 \times 10^{-2}$ \unit{\per\micro\second}.
We show the stationary distribution and empirical committor projected onto the CVs in Fig.~\ref{fig:trpcage}.
The unfolded state $A$ is in the upper right, and the folded state $B$ is in the lower left.

We use as inputs to the neural networks three sets of features: the three RMSDs used to define the states (``RMSDs''), sines and cosines of the $(\phi,\psi)$ backbone dihedral angles (``Dihedrals''), and all pairwise distances between C$_\alpha$ atoms (``Distances'').

\subsubsection{Villin}

The fast-folding 35-residue villin headpiece subdomain, hereafter referred to as villin, is one of the most well-studied protein folding models both experimentally and computationally (see Ref.~\citenum{wang_novel_2019} and references therein).
In the folded state, its secondary structure consists of three $\alpha$-helices spanning residues 3--10, 14--19, and 22--32; these helices pack around a hydrophobic core centered on residues Phe6, Phe10, and Phe17.
We study the K65nL/N68H/K70nL mutant (sequence: \seqsplit{LSDEDFKAVFGMTRSAFANLPLW{nL}QQHL{nL}KEKGLF}, where nL is norleucine), which is engineered to fold more rapidly.
The dataset consists of a single \qty{125}{\micro\second} trajectory, saved every \qty{0.2}{\nano\second} \cite{lindorff-larsen_how_2011}. To generate training and validation data, the trajectory is split in half and treated as two independent segments.

\begin{figure*}
  \includegraphics[scale=0.9]{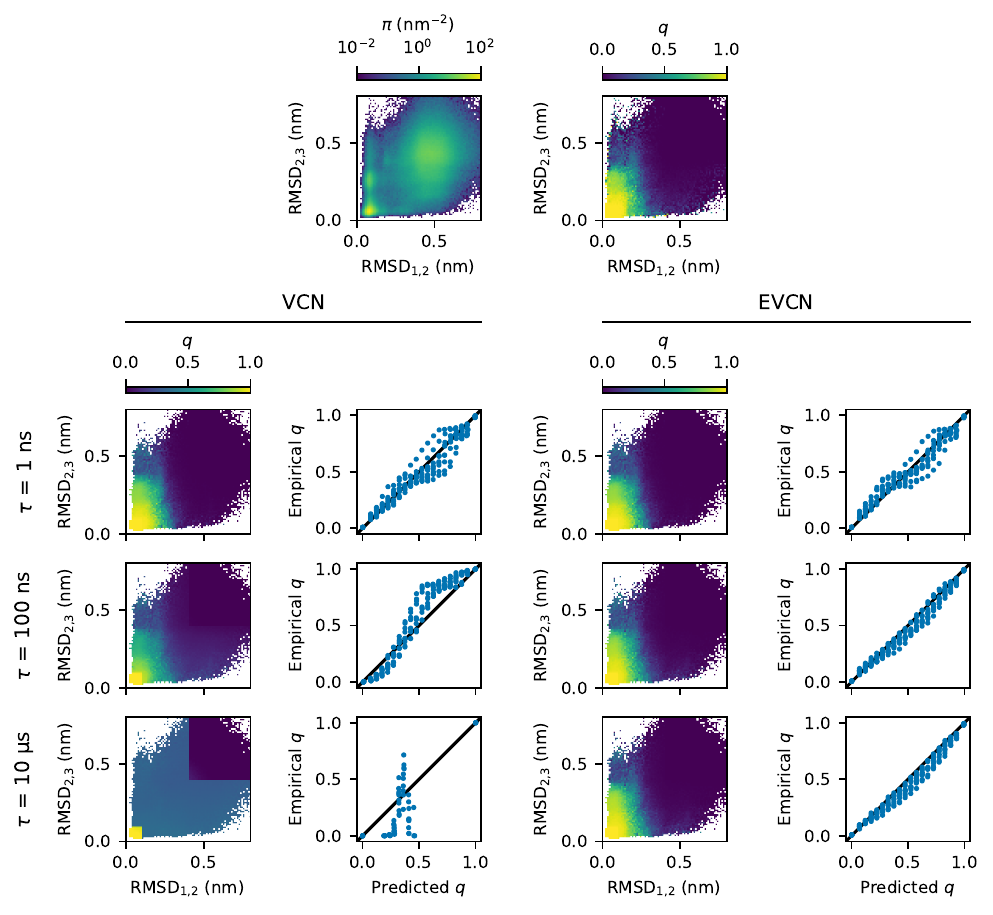}
  \caption{
    CV projections and reliability diagrams for villin.
    (top left) CV projection of the stationary distribution.
    (top right) CV projection of the empirical committor.
    (left columns of VCN and EVCN) CV projections of the predicted committors.
    (right columns of VCN and EVCN) Reliability diagrams for the predicted committors.
    All results shown are obtained with the dihedral neural-network inputs.
  }
  \label{fig:villin}
\end{figure*}

We define three CVs: the RMSD of the C$_\alpha$ atoms to the experimental structure (PDB 2F4K \cite{kubelka_sub-microsecond_2006}) of helices 1 and 2 ($\CaRMSD{1,2}$), helices 2 and 3 ($\CaRMSD{2,3}$), and helices 1, 2, and 3 ($\CaRMSD{1,2,3}$).
The unfolded state $A$ is defined as configurations with $\CaRMSD{1,2} \ge \qty{0.4}{\nano\meter}$, $\CaRMSD{2,3} \ge \qty{0.4}{\nano\meter}$, and $\CaRMSD{1,2,3} \ge \qty{0.5}{\nano\meter}$, and the folded state $B$ is defined as configurations with $\CaRMSD{1,2} \le \qty{0.1}{\nano\meter}$, $\CaRMSD{2,3} \le \qty{0.1}{\nano\meter}$ and $\CaRMSD{1,2,3} \le \qty{0.15}{\nano\meter}$.
With these state definitions, unfolded to folded transitions occur at a rate of 
 $3.7 \times 10^{-1}$ \unit{\per\micro\second}.
We show the stationary distribution and empirical committor projected onto $\CaRMSD{1,2}$ and $\CaRMSD{2,3}$ in Fig.~\ref{fig:villin}.
The unfolded state $A$ is in the upper right, and the folded state $B$ is in the lower left.

We use as inputs to the neural networks three sets of features: the three RMSDs (``RMSDs''), sines and cosines of the $(\phi,\psi)$ backbone dihedral angles (``Dihedrals''), and all pairwise distances between C$_\alpha$ atoms (``Distances'').


\clearpage

\subsection{Committor}\label{sec:committor_results}

We examine the behavior of committors predicted using the VCN and EVCN loss functions at different lag times.
We compute committors for individual conformations in the sampled trajectories and then visualize the results in two ways.  First, we plot the average committor for structures in a bin defined by a pair of CV values.  Second, we partition the predicted committor into 20 uniformly spaced bins over $[0,1]$, and for each bin, we plot the mean empirical committor against the mean predicted committor.  In the latter set of plots, which we term ``reliability diagrams'' in line with the machine learning literature (which also calls them ``calibration curves''), each point is from a single model.
Perfect alignment between the predicted and empirical committors corresponds to points lying along the diagonal (shown in black).

We first look at AIB\textsubscript{9} (Fig.~\ref{fig:aib9}).
At the shortest lag time (\qty{1}{\pico\second}), CV projections of the predicted committor exhibit a checkered pattern reflecting the boundaries of the intermediate states.
The reliability diagrams show substantial noise, but the means follow the diagonal.
These artifacts arise because the VCN and EVCN losses are both relatively insensitive to slowly decaying modes, encouraging the committor to be flat within intermediate states at the expense of transition states between them.
At the intermediate lag time (\qty{100}{\pico\second}), the committor becomes qualitatively accurate for both loss functions.
At the longest lag time (\qty{10}{\nano\second}), the committor for EVCN remains accurate, while the committor for VCN appears flattened in the transition region and exhibits large jumps near the boundaries of states $A$ and $B$.
These behaviors are reflected in the reliability diagrams;  there are significant deviations from the diagonal for VCN but not EVCN.
We can explain this behavior by considering $\E[\hat{u}(X_0) \hat{u}(X_\tau)] \approx \E[\hat{u}(X_0)] \E[\hat{u}(X_\tau)]$ as $\tau \to \infty$.
With this approximation, \eqref{eq:vcn_loss} is minimized when $u(x) = \1_B(x)$ for $x \notin D$ and $u(x) = \E[\1_B(X_0)] / \E[\1_{D^\comp}(X_0)]$ otherwise.

Next, we look at Trp-cage (Fig.~\ref{fig:trpcage}).
At the \qty{1}{\nano\second} lag time, both loss functions incorrectly predict the committor of the intermediate state above the folded basin in the CV projection to be near zero.
Reflecting this, the reliability diagram has most points above the diagonal, where the empirical committor is greater than the predicted committor.
At the \qty{100}{\nano\second} lag time, both methods resolve the intermediate state, but the VCN predictions are lower (darker CV projection). Correspondingly, the reliability diagram for VCN has more points above the diagonal. For EVCN, on the other hand, the points are noisy but do not show obvious bias.
With a \qty{10}{\micro\second} lag time, VCN breaks down.
The CV projection shows that the predicted committor is flat in $D$, and the reliability diagram shows that all the predicted values (other than 0 in $A$ and 1 in $B$) are clustered near a single value.
On the other hand, the CV projection of the EVCN result remains accurate, and the reliability diagram follows the diagonal with less noise than the \qty{100}{\nano\second} lag time.

Villin (Fig.~\ref{fig:villin}) exhibits similar behavior to the previous systems.
For the \qty{1}{\nano\second} lag time, both loss functions yield similar, good predictions.
However, at the \qty{100}{\nano\second} lag time, the EVCN prediction shows clear improvement over the VCN prediction, which is already beginning to flatten in the transition region.
In the reliability diagrams, EVCN predictions are near the diagonal, while VCN predictions visibly deviate.
At the \qty{10}{\micro\second} lag time, the VCN prediction is again clustered around a single value, while the EVCN prediction remains accurate.

\begin{figure*}
\includegraphics[scale=0.9]{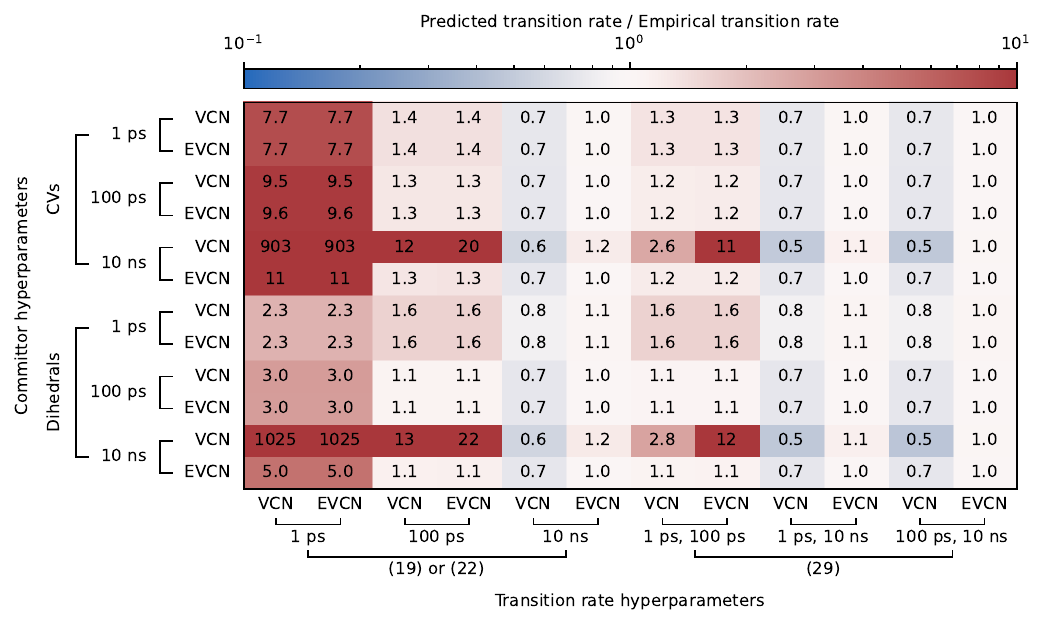}
\caption{
 Ratio of predicted to empirical transition rates for AIB\textsubscript{9}, for different committor and transition rate estimator hyperparameters.
The empirical transition rate  is $9.1 \times 10^{-3}$ \unit{\per\nano\second}.
\label{fig:aib9_rate}
}
\end{figure*}

\begin{figure*}
\includegraphics[scale=0.9]{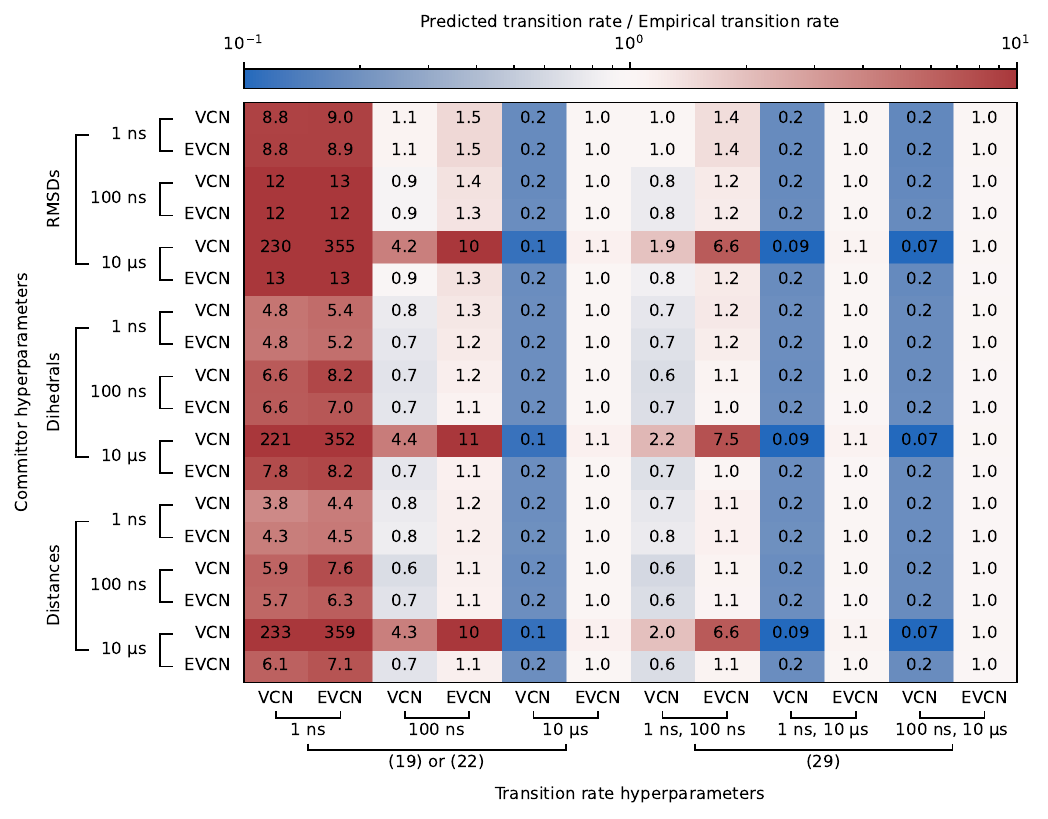}
\caption{
 Ratio of predicted to empirical transition rates for Trp-cage, for different committor and transition rate estimator hyperparameters.
The empirical transition rate  is $7.7 \times 10^{-2}$ \unit{\per\micro\second}.
\label{fig:trpcage_rate}
}
\end{figure*}

\begin{figure*}
\includegraphics[scale=0.9]{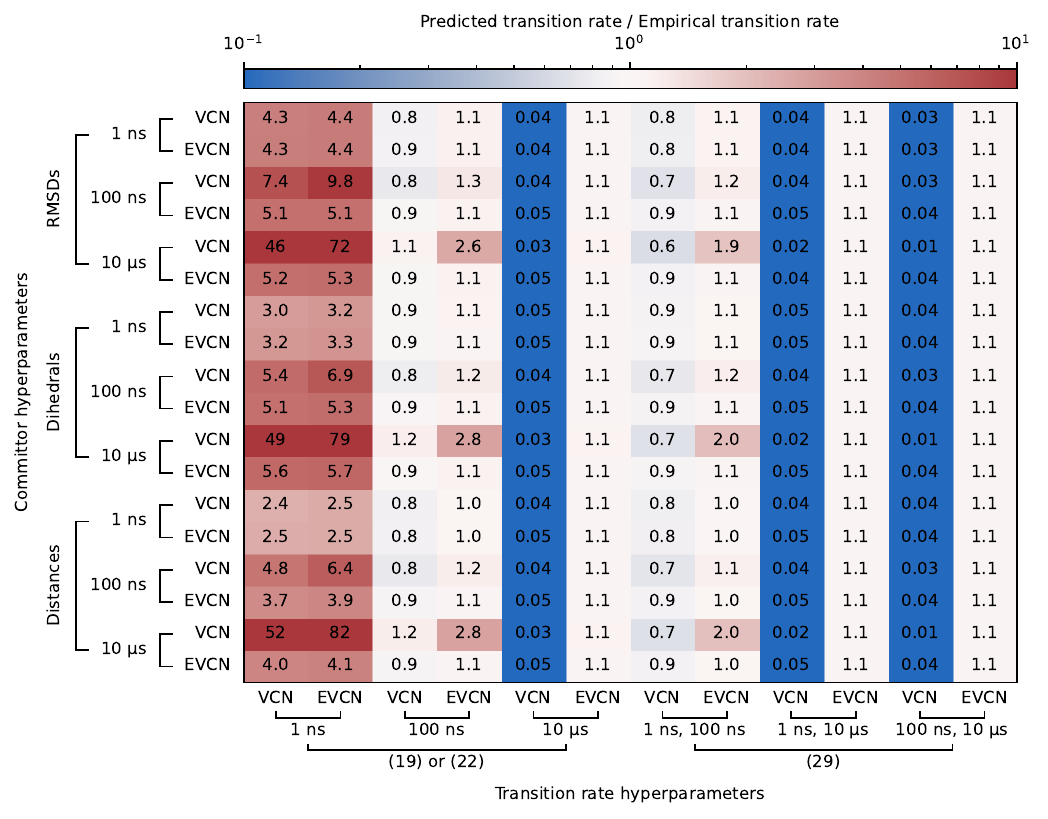}
\caption{
 Ratio of predicted to empirical transition rates for villin, for different committor and transition rate estimator hyperparameters.
The empirical transition rate  is $3.7 \times 10^{-1}$ \unit{\per\micro\second}.
\label{fig:villin_rate}
}
\end{figure*}

\subsection{Transition rate}


In  Figs.~\ref{fig:aib9_rate}--\ref{fig:villin_rate}, we examine the transition rates predicted by the VCN and EVCN loss functions with various lag times.  The loss function and lag time can affect the transition rates through both training the neural network for the committor and application of the rate formula given the trained committor.  To distinguish these effects, we evaluate the rates using loss functions and lag times (columns) that are independent of those used to obtain the committors (rows).
Unsurprisingly, more accurate committors yield more accurate transition rates  (the large errors in the committors for VCN at long lag times discussed in the previous section appear as horizontal stripes of darker colors in Figs.~\ref{fig:aib9_rate}--\ref{fig:villin_rate}).
Still, even for a given committor (row), the loss function and lag time used to predict the rate (column) noticeably affect the result.
As in the previous section, rates are overestimated at short lag times  (red in Figs.~\ref{fig:aib9_rate}--\ref{fig:villin_rate}) and approach their correct values as lag times increase  (pale colors). At long lag times, the VCN loss predicts rates that go to zero (approximately proportional to $1/\tau$, as can be seen by substituting a constant committor value for $x\in D$, consistent with the flattening observed above;  blue in Figs.~\ref{fig:aib9_rate}--\ref{fig:villin_rate}), whereas the EVCN loss predicts rates that plateau near the true value  (pale colors).
The choice of input features has a larger impact at shorter lag times.
Less expressive input features---such as the CVs for AIB\textsubscript{9} and the RMSDs for Trp-cage and villin---tend to produce larger overestimates of the transition rate  (darker red).

We examine the estimator in \eqref{eq:approx_rate} in  the right six columns of Figs.~\ref{fig:aib9_rate}--\ref{fig:villin_rate}.
We observe that using \eqref{eq:approx_rate} with EVCN consistently yields more accurate rate estimates  (paler colors) than \eqref{eq:evcn_loss} by itself.
Rates predicted using the VCN loss \eqref{eq:vcn_loss} or its counterpart with \eqref{eq:approx_rate} are not variational and tend to underestimate the rate  (blue in Figs.~\ref{fig:aib9_rate}--\ref{fig:villin_rate}) at intermediate and long lag times.
For both VCN and EVCN, the choice of the longer of the two lag times in \eqref{eq:approx_rate} appears to matter more than the choice of the shorter of the two lag times.

\begin{figure*}
\includegraphics{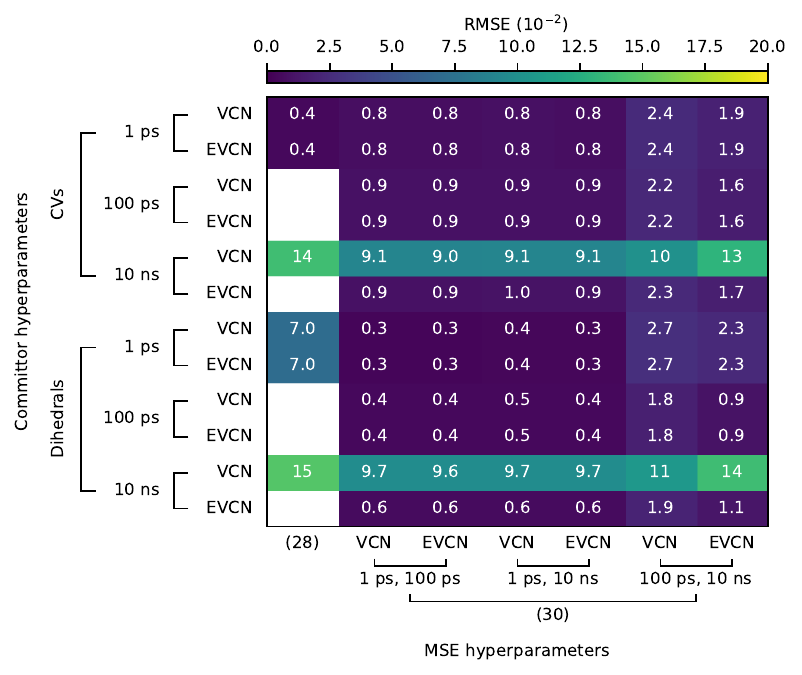}
\caption{
 RMSE in the predicted committor for AIB\textsubscript{9}, for different committor and MSE estimator hyperparameters.
MSE estimates with negative values are omitted.
}
\label{fig:aib9_rmse}
\end{figure*}

\begin{figure*}
\includegraphics{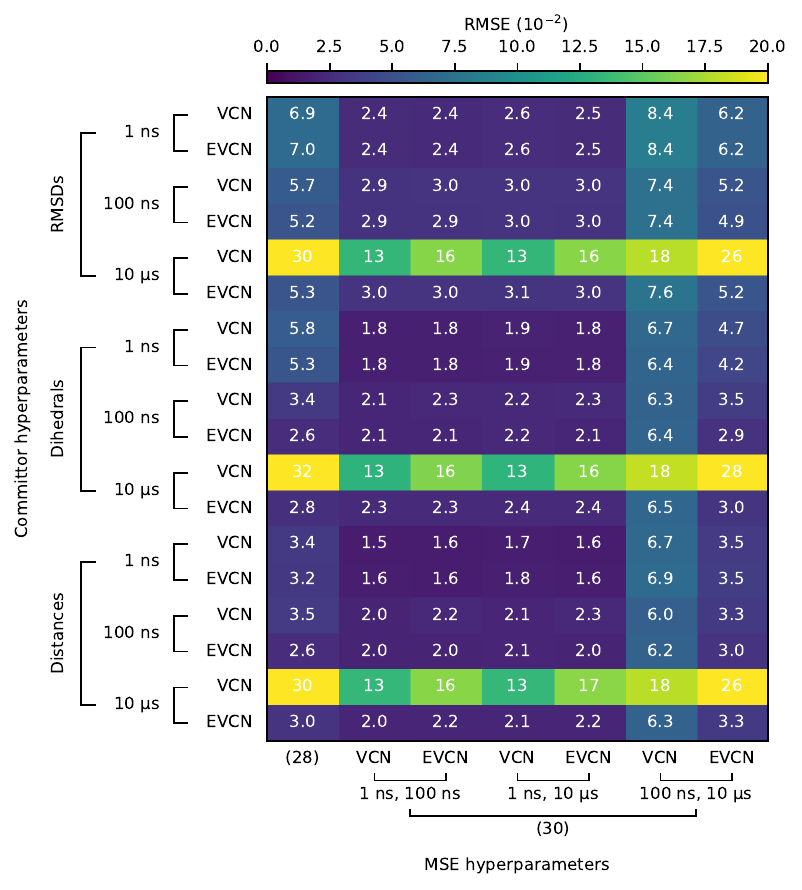}
\caption{
 RMSE in the predicted committor for Trp-cage, for different committor and MSE estimator hyperparameters.
}
\label{fig:trpcage_rmse}
\end{figure*}

\begin{figure*}
\includegraphics{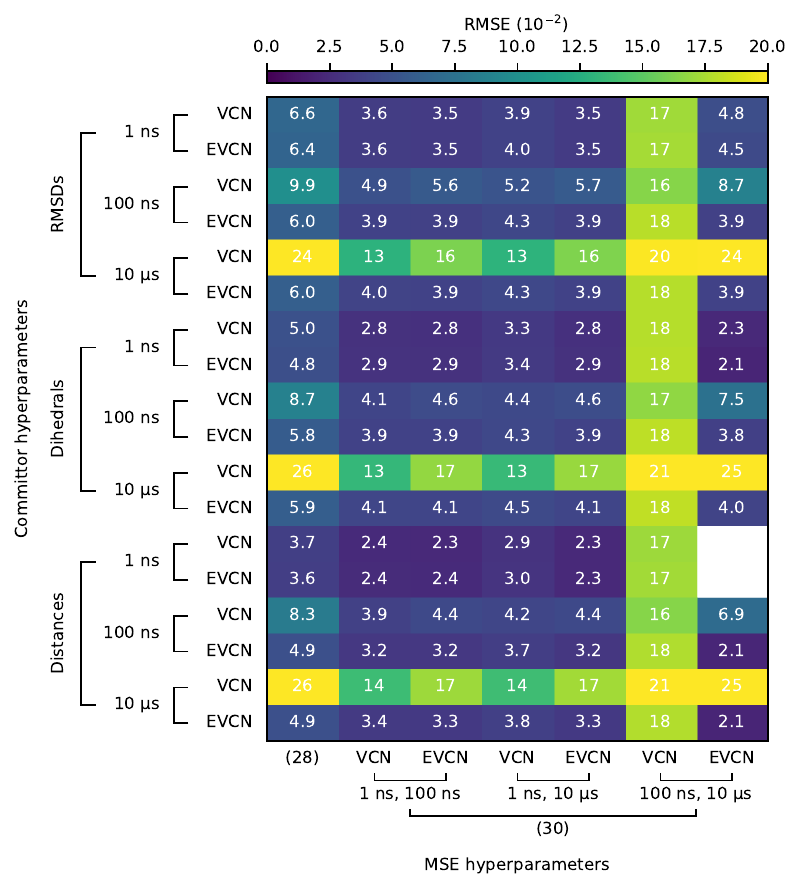}
\caption{
 RMSE in the predicted committor for villin, for different committor and MSE estimator hyperparameters.
MSE estimates with negative values are omitted.
}
\label{fig:villin_rmse}
\end{figure*}

\subsection{Mean squared error}\label{sec:mse_results}

In  Figs.~\ref{fig:aib9_rmse}--\ref{fig:villin_rmse}, we show root MSE (RMSE) estimates (mean of 10 models) for the predicted committor obtained with various training hyperparameters.  
Focusing first on the  leftmost column computed with \eqref{eq:mse_2_traj}, at short lag times, committors trained with the VCN and EVCN loss function have similar errors.
Notably, more expressive features (dihedral angles and distances) do not always yield lower error.
As the lag time increases, the VCN RMSE values increase markedly, while the EVCN RMSE values remain stable.

In  Figs.~\ref{fig:aib9_rmse}--\ref{fig:villin_rmse}, we also compare \eqref{eq:approx_mse} with \eqref{eq:mse_2_traj}.  The two estimators generally track each other, but \eqref{eq:approx_mse} can differ from \eqref{eq:mse_2_traj} by an order of magnitude when using VCN with long lag times  (or about a factor of three in Figs.~\ref{fig:aib9_rmse}--\ref{fig:villin_rmse}, which show the RMSE).  For villin, the approximation becomes negative for ECVN at long lag times  (preventing us from taking the square root), likely due to statistical noise.
In contrast to \eqref{eq:approx_rate}, for a given committor (row), the choice of the shorter of the two lag times appears to have more  impact on the MSE approximated by \eqref{eq:approx_mse} than the choice of the longer of the two lag times.


\subsection{Summary}

In summary, VCN committors tend to flatten at long lag times, while EVCN committors remain accurate.
For the transition rates, we considered the effects of varying the committor hyperparameters and the rate hyperparameters separately.
When varying the committor hyperparameters, VCN committors underestimate the rate and have high MSE, compared to EVCN committors. When varying the rate hyperparameters, both methods are similarly bad at short lag times, but EVCN converges to the correct rate at long lag times, whereas VCN underestimates it.
The estimator based on two lag times, when used with ECVN, provides the most accurate transition rates, particularly for intermediate lag times.

\section{Discussion \& Conclusions}\label{sec:discussion}

In this work, we derived an exact multiple-time-step expression for the transition rate and demonstrated, using models of molecular systems, that a variational scheme based on this expression yields accurate committors and transition rates with relatively little sensitivity to the choice of lag time.
Our approach extends previous methods \cite{chen_discovering_2023}, which are valid only in the limit of a single-time-step lag time and break down at longer lag times.
The key idea is that trajectories must be ``stopped'' at the boundaries (i.e., when they first enter $A$ and $B$) \cite{strahan_long-time-scale_2021,tuchkov_error_2025} and these stopping times used in the variational expression; these considerations furthermore give rise naturally to terms in the loss that penalize violation of the boundary conditions.

The datasets that we used for our numerical tests \cite{strahan_inexact_2023,lindorff-larsen_how_2011} consist of long, equilibrium trajectories, which spend most of their time in the reactant and product states. Theoretical~\cite{cheng_surprising_2024} and empirical~\cite{strahan_long-time-scale_2021,strahan_predicting_2023} evidence indicates that data distributed according to the equilibrium distribution are not sufficient to learn rare-event statistics accurately.
This may explain why standard regularization techniques were unable to address the overfitting we experienced.
In the future, it would be worth exploring integration over lag times, as in Ref.~\citenum{lorpaiboon_integrated_2020}, and regularization strategies that directly restrain the committor to depend on physical or learned CVs in a simple way.

Better control of sampling and, in particular, sampling of key transition pathways can be achieved by datasets that consist of short trajectories \cite{strahan_long-time-scale_2021,strahan_inexact_2023,cheng_surprising_2024}. Short trajectories have the disadvantage that longer lag times are inaccessible, making it harder to capture information about slow modes, but this issue can be partially addressed by using loss functions with  sensitivity to these modes  even at intermediate lag times, such as \eqref{eq:approx_rate}, introduced here.
More advanced approaches incorporating memory, such as those proposed in Ref.~\citenum{liu_memory_2025}, may further improve accuracy.
However, those approaches do not overcome the fundamental issue that both VCN and EVCN require samples with equilibrium weighting, which, as discussed above emphasizes the stable states at the expense of the transition pathways.

The schemes proposed in Ref.~\citenum{strahan_predicting_2023} and  Ref.~\citenum{strahan_inexact_2023} allow both arbitrary non-equilibrium data distributions and irreversible dynamics, and have shown promise on several benchmark problems. However, these schemes have tradeoffs relative to the variational one proposed in this article.  
The method proposed in Ref.~\citenum{strahan_predicting_2023} requires a trajectory dataset with multiple short forward simulations for every initial configuration, and the method proposed in Ref.~\citenum{strahan_inexact_2023} uses a training procedure that is not the gradient of any loss function when the data are non-equilibrium or the dynamics are irreversible, which can complicate training.
Developing a method that combines the strengths of the  short-trajectory approaches with the variational ones remains an open problem.

\begin{appendices}

\section{Derivation of the transition rate}
\label{sec:tpt_rate}
In this subsection, we use transition path theory to derive expressions for the transition rate in terms of committors and finite-length trajectories.
In brief, we first define the transition rate ($\Phi$) in terms of the number of transitions within a time interval, then express $\Phi$ as an average of changes in a reaction coordinate over single and multiple time steps.

We denote discrete time intervals by $[t,t'] = \{t,t+\dt,\dots,t'\}$. A transition path is a trajectory segment $X_{[s,s']} = (X_s,X_{s+\dt},\dots,X_{s'})$ with $X_s \in A$, $X_{s'} \in B$, and $X_t \in D$ for $s < t < s'$.
We define
\begin{equation}
  H_{AB}(X_{[s,s']}) = \1_A(X_s) \prod_{t \in [s+\dt,s-\dt]} \1_D(X_t) \1_B(X_{s'}),
\end{equation}
which is 1 if $X_{[s,s']}$ is a transition path and 0 otherwise.
The number of transition paths from $A$ to $B$ within the trajectory segment $X_{[t,t']}$ is
\begin{equation}
  N_{AB}(X_{[t,t']}) = \sum_{s \in [t,t'-\dt]} \sum_{s' \in [s+\dt,t']} H_{AB}(X_{[s,s']}),
  \label{eq:num_tp}
\end{equation}
and the transition rate $\Phi$ is the mean of $N_{AB}$ per time:
\begin{equation}
  \Phi = \lim_{T \to \infty} \frac{1}{2 T} N_{AB}(X_{[-T,T]}).
  \label{eq:rate_def}
\end{equation}

We now express \eqref{eq:rate_def} in terms of a reaction coordinate $\xi$ that satisfies $\xi(x) = 0$ for $x \in A$ and $\xi(x) = 1$ for $x \in B$.
We start with the identity
\begin{equation}
  H_{AB}(X_{[s,s']})
  = H_{AB}(X_{[s,s']}) \sum_{t \in [s,s'-\dt]} (\xi(X_{t+\dt}) - \xi(X_t)),
\end{equation}
which states that the total change in progress along a transition path is one.
Mathematically, the sum telescopes, and $\xi(X_{s'}) - \xi(X_s) = 1$ for transition paths because $X_s \in A$ and $X_{s'} \in B$.
We substitute this identity into the summand of \eqref{eq:num_tp} and interchange the sums, yielding
\begin{equation}
  N_{AB}(X_{[t,t']})
  =
  \sum_{t'' \in [t,t'-\dt]}
  \sum_{s \in [t,t'']}
  \sum_{s' \in [t''+\dt,t']}
  H_{AB}(X_{[s,s']})
  (\xi(X_{t''+\dt}) - \xi(X_{t''})).
  \label{eq:num_tp_rearrange}
\end{equation}
We now set $t=-T$ and $t'=T$ and let $T \to \infty$; \eqref{eq:rate_def} becomes
\begin{equation} \label{eq:rate_traj}
  \Phi = \lim_{T \to \infty} \frac{1}{2 T} \sum_{t \in [-T,T-\dt]} F_{t,t+\dt},
\end{equation}
where $F_{t,t+\dt}$ is the reaction progress from time $t$ to time $t+\dt$:
\begin{equation}
  F_{t,t+\dt}
  =
  \sum_{s \in [-\infty,t]}
  \sum_{s' \in [t+\dt,\infty]}
  H_{AB}(X_{[s,s']})
  (\xi(X_{t+\dt}) - \xi(X_t)).
  \label{eq:progress_def_1}
\end{equation}


The standard transition path theory expression for the transition rate uses single-step trajectories.
In that case, the reaction progress \eqref{eq:progress_def_1} can be expressed as
\begin{equation}
  F_{t,t+\dt} = \1_A(X_{\bar{S}_t}) \1_B(X_{S_{t+\dt}}) (\xi(X_{t+\dt}) - \xi(X_t)),
  \label{eq:progress_1}
\end{equation}
using the identities
\begin{align}
  \1_A(X_{\bar{S}_t}) & = \sum_{s \in [-\infty,t]} \1_A(X_s) \prod_{t' \in [s+\dt,t]} \1_D(X_{t'}),
  \label{eq:prev_a} \\
  \1_B(X_{S_t}) & = \sum_{s \in [t,\infty]} \prod_{t' \in [t,s-\dt]} \1_D(X_{t'}) \1_B(X_s).
  \label{eq:next_b}
\end{align}
Using \eqref{eq:qp_def} and \eqref{eq:qm_def}, the expectation of \eqref{eq:progress_1} conditioned on single-step trajectories $X_{[t,t+\dt]} = (X_t,X_{t+\dt})$, denoted by $f(X_{[t,t+\dt]})$, is
\begin{equation}
  f(X_{[t,t+\dt]}) = \bar{q}(X_t) q(X_{t+\dt}) (\xi(X_{t+\dt}) - \xi(X_t)).
\end{equation}
We can then use ergodicity to express \eqref{eq:rate_traj} as
\begin{equation} \label{eq:tpt_rate_1_app}
  \Phi = \frac{1}{\dt} \E[f(X_{[0,\dt]})],
\end{equation}
where the expectation is over paths sampled at equilibrium.

In previous work, we proposed and applied a multiple-step expression for the transition rate. \cite{strahan_long-time-scale_2021,antoszewski_kinetics_2021}
There, we applied \eqref{eq:qp_bvp} and \eqref{eq:qm_bvp} to an average of \eqref{eq:tpt_rate_1_app}:
\begin{align}
  \Phi
  & = \frac{1}{\tau} \sum_{t \in [0,\tau-\dt]} \E[f(X_{[t,t+\dt]})] \\
  & = \frac{1}{\tau}
  \sum_{t \in [0,\tau-\dt]}
  \E[
    \bar{q}(X_{0 \bmax \bar{S}_{t}})
    q(X_{\tau \bmin S_{t+\dt}})
    (\xi(X_{t+\dt}) - \xi(X_{t}))
  ].
\end{align}
In this work, we derive an equivalent expression that depends on committors only at the endpoints.
To this end, we note that the reaction progress from time $t$ to time $t'$ can be written as a sum over single steps:
\begin{equation}
  F_{t,t'} = \sum_{t'' \in [t,t'-\dt]} F_{t'',t''+\dt}.
  \label{eq:progress_def}
\end{equation}
We then substitute \eqref{eq:progress_def_1} into \eqref{eq:progress_def}, split the sums by whether the transition path starts before time $t$ and ends after time $t'$, and interchange and telescope the sums. The result is
\begin{align}
  F_{t,t'}
  & = \sum_{s \in [-\infty,t-\dt]} \sum_{s' \in [t'+\dt,\infty]} H_{AB}(X_{[s,s']}) (\xi(X_{t'}) - \xi(X_t)) \nonumber \\
  & \peq {} + \sum_{s \in [t,t']} \sum_{s' \in [t'+\dt,\infty]} H_{AB}(X_{[s,s']}) (\xi(X_{t'}) - 0) \nonumber \\
  & \peq {} + \sum_{s \in [-\infty,t-\dt]} \sum_{s' \in [t,t']} H_{AB}(X_{[s,s']}) (1 - \xi(X_t)) \nonumber \\
  & \peq {} + \sum_{s \in [t,t'-\dt]} \sum_{s' \in [s+\dt,t']} H_{AB}(X_{[s,s']}) (1 - 0) \\
  & = \1_A(X_{\bar{S}_t}) \1_D(X_{t' \bmin S_t}) \1_B(X_{S_{t'}}) (\xi(X_{t'}) - \xi(X_t)) \nonumber \\
  & \peq {} + \1_A(X_{t \bmax \bar{S}_{t'}}) \1_B(X_{S_{t'}}) (\xi(X_t) - 0) \nonumber \\
  & \peq {} + \1_A(X_{\bar{S}_t}) \1_B(X_{t' \bmin S_t}) (1 - \xi(X_t)) \nonumber \\
  & \peq {} + N_{AB}(X_{[t,t']}),
\end{align}
where we have used \eqref{eq:prev_a}, \eqref{eq:next_b}, and
\begin{align}
  \1_A(X_{t \bmax \bar{S}_{t'}}) & = \sum_{s \in [t,t']} \1_A(X_s) \prod_{t'' \in [s+\dt,t']} \1_D(X_{t''}), \\
  \1_B(X_{t' \bmin S_t}) & = \sum_{s \in [t,t']} \prod_{t'' \in [t,s-\dt]} \1_D(X_{t''}) \1_B(X_s), \\
  \1_D(X_{t' \bmin S_t}) & = \1_D(X_{t \bmax \bar{S}_{t'}}) = \prod_{t'' \in [t,t']} \1_D(X_{t''}).
\end{align}
For times within the interval $[t,t']$, these indicator functions identify  whether the system last exited $A$ before time $t'$, first entered $B$ after time $t$, or was always in $D$, respectively.
The expectation of $F_{t,t'}$ conditioned on multiple-step trajectories $X_{[t,t']}$, denoted by $f(X_{[t,t']})$, is
\begin{align}
  f(X_{[t,t']}) = {}
  & \bar{q}(X_t) \1_D(X_{t' \bmin S_t}) q(X_{t'}) (\xi(X_{t'}) - \xi(X_t))
  \nonumber \\
  & + \1_A(X_{t \bmax \bar{S}_{t'}}) q(X_{t'}) (\xi(X_{t'}) - 0)
  \nonumber \\
  & + \bar{q}(X_t) \1_B(X_{t' \bmin S_t}) (1 - \xi(X_t))
  \nonumber \\
  & + N_{AB}(X_{[t,t']}),
  \label{eq:tpt_rate_telescope_app}
\end{align}
where we have used \eqref{eq:qp_def} and \eqref{eq:qm_def}.
By ergodicity, \eqref{eq:rate_traj} is
\begin{equation} \label{eq:tpt_rate_app}
  \Phi = \frac{1}{\tau} \E[f(X_{[0,\tau]})].
\end{equation}

The key point is that transition paths observed in the time interval $[0,\tau]$ can be classified into four cases by their starting and ending times: starting before $0$ and ending after $\tau$, starting before $0$ and ending within $[0,\tau]$, starting within $[0,\tau]$ and ending after $\tau$, and both starting and ending within $[0,\tau]$.
Each term in \eqref{eq:tpt_rate_telescope} corresponds to one of these cases and represents the product of the probability of observing that type of transition path and the progress it contributes within the time interval $[0,\tau]$.

\section{Mode decomposition of the variational loss function}
\label{sec:modedecomp}

When detailed balance is satisfied, we can write a perturbation $\eta$ to $q$ as in \eqref{eq:evcn_perturb} as a sum over eigenvalues $\{e^{-\tau/\sigma_k}\}$ and eigenfunctions $\{v_k\}$:
\begin{equation}
  \E[\1_D(X_{\tau \bmin S_0}) \eta(X_\tau) \mid X_0 = x]
  = \sum_k e^{-\tau/\sigma_k} v_k(x) \langle v_k , \eta \rangle,
\end{equation}
where $\tau \ge 0$,
$\langle v_k, \eta \rangle = \int v_k(x) \eta(x) \pi(x) \odif{x}$,
$\langle v_k , v_l \rangle = \delta_{kl}$,
and $\delta_{kl}$ is the Kronecker delta.
The estimated transition rate is then
\begin{alignat}{2}
  \tilde{L}_\tau(q + \eta)
  & = \Phi + \frac{1}{2\tau} \E[\eta(X_0)^2 + \eta(X_\tau)^2 - 2 \1_D(X_{\tau \bmin S_0}) \eta(X_0) \eta(X_\tau)]
  \\
  & = \Phi + \sum_k \frac{1 - e^{-\tau / \sigma_k}}{\tau} \langle v_k , \eta \rangle^2.
  \label{eq:mode_expansion}
\end{alignat}
Since $\langle v_k , \eta \rangle^2 \ge 0$ and $(1 - e^{-\tau/\sigma_k}) / \tau$ is a nonnegative, monotonically decreasing function of $\tau$, the estimated transition rate decreases (and accuracy increases) with increasing $\tau$ for fixed $\eta$.
For $\sigma_k \ll \tau$, $(1 - e^{-\tau/\sigma_k}) / \tau \approx 1/\tau$, while for $\sigma_k \gg \tau$, $(1 - e^{-\tau/\sigma_k}) / \tau \approx 1/\sigma_k$.
Thus, the loss is similarly sensitive to components of $\eta$ with decay times shorter than $\tau$, and progressively less sensitive to those with longer decay times.

\section{Mode decomposition of the mean squared error and alternative transition rate expression}

\label{sec:Lextrapolation}

As shown by the mode expansion in \eqref{eq:mode_expansion}, transition rate estimates at short lag times tend to overestimate the true rate and can be improved by extrapolating the rate to infinite lag time.
We do so linearly using the points $(1/\tau_1,\tilde{L}_{\tau_1}(u))$ and $(1/\tau_2,\tilde{L}_{\tau_2}(u))$:
\begin{align}
  \tilde{L}_{\tau_1,\tau_2}(u)
  & = \tilde{L}_{\tau_1}(u) + \frac{\tilde{L}_{\tau_2}(u) - \tilde{L}_{\tau_1}(u)}{1/\tau_2 - 1/\tau_1} \Bigl( \frac{1}{\infty} - \frac{1}{\tau_1} \Bigr)
  \\
  & = \frac{\tau_1 \tilde{L}_{\tau_1}(u) - \tau_2 \tilde{L}_{\tau_2}(u)}{\tau_1 - \tau_2}.
  \label{eq:approx_rate_app}
\end{align}
Surprisingly, \eqref{eq:approx_rate_app} is variational: it satisfies $\tilde{L}_{\tau_1,\tau_2}(u) \ge \Phi$ for all $\tau_1,\tau_2 \ge 0$.
Using the mode expansion in \eqref{eq:mode_expansion}, we find
\begin{equation}
  \tilde{L}_{\tau_1,\tau_2}(q+\eta) = \Phi + \sum_k \frac{e^{-\tau_1/\sigma_k} - e^{-\tau_2/\sigma_k}}{\tau_2 - \tau_1} \langle v_k , \eta \rangle^2,
\end{equation}
in which each coefficient of $\langle v_k , \eta \rangle^2$ is nonnegative.
By the same approach, it can be shown that \eqref{eq:approx_rate_app} is, in fact, the minimum variational affine combination of $\tilde{L}_{\tau_1}$ and $\tilde{L}_{\tau_2}$.
To the best of our knowledge, the variational expression in \eqref{eq:approx_rate_app} has not been proposed previously.

We now examine how $\tilde{L}_{\tau_1,\tau_2}(u)$ varies with $\tau_1$ and $\tau_2$.
It is symmetric ($\tilde{L}_{\tau_1,\tau_2}(u) = \tilde{L}_{\tau_2,\tau_1}(u)$), decreases with increasing $\tau_1$ or $\tau_2$, and is bounded by
\begin{equation}
  \tilde{L}_{\tau_2}(u) = \tilde{L}_{0,\tau_2} \ge \tilde{L}_{\tau_1,\tau_2}(u) \ge \tilde{L}_{\infty,\tau_2}(u) = \Phi.
\end{equation}
To understand the behavior away from these bounds, we consider a representative case $\tilde{L}_{\tau,\tau}(u) = \odv{(\tau \tilde{L}_\tau(u))}/{\tau}$, which has a simple mode expansion:
\begin{equation}
  \tilde{L}_{\tau,\tau}(q+\eta) = \Phi + \sum_k \frac{e^{-\tau/\sigma_k}}{\sigma_k} \langle v_k , \eta \rangle^2.
\end{equation}
As $\tau$ increases, each coefficient of $\langle v_k , \eta \rangle^2$ exponentially decays from $1/\sigma_k$ to zero.
Unlike $\tilde{L}_\tau$, $\tilde{L}_{\tau,\tau}$ is not monotonic in $\sigma_k$: the coefficient is zero at both $\sigma_k = 0$ and $\sigma_k = \infty$, and has a maximum of $1 / (\tau e)$ at $\sigma_k = \tau$.

Similarly, we can use the mode expansion in \eqref{eq:mode_expansion} to understand the associated approximation of the MSE in \eqref{eq:approx_mse}.  To this end, we write
\begin{align}
  \MSE_{\tau_1,\tau_2}(q+\eta) = \MSE(q+\eta) - \sum_k \frac{e^{-\tau_1/\sigma_k}/\tau_1-e^{-\tau_2/\sigma_k}/\tau_2}{1/\tau_1-1/\tau_2} \langle v_k , \eta \rangle^2.
\end{align}
Note the mode expansion of the MSE is
\begin{equation}
  \MSE(q + \eta) = \sum_k \langle v_k , \eta \rangle^2.
\end{equation}
Because $0 \le (e^{-\tau_1/\sigma_k}/\tau_1-e^{-\tau_2/\sigma_k}/\tau_2)/(1/\tau_1-1/\tau_2) \le 1$, this expression always underestimates the MSE in the limit of infinite data.
It can be shown that \eqref{eq:approx_mse} is the maximum linear combination of $\tilde{L}_{\tau_1}$ and $\tilde{L}_{\tau_2}$ that is a lower bound for the MSE.
More precisely, $\MSE_{\tau_1,\tau_2}$ is symmetric ($\MSE_{\tau_1,\tau_2}(u) = \MSE_{\tau_2,\tau_1}(u)$), increases with increasing $\tau_1$ or $\tau_2$, and is bounded by
\begin{equation}
  0 = \MSE_{0,\tau_2}(u) \le \MSE_{\tau_1,\tau_2}(u) \le \MSE_{\infty,\tau_2}(u) = \tau_2 (\tilde{L}_{\tau_2}(u) - \Phi) \le \MSE(u).
\end{equation}
For $\tau_1$ close to $\tau_2$, rather than near $0$ or $\infty$, we consider a representative case $\MSE_{\tau,\tau}(u) = \odv{\tilde{L}_\tau(u)}/{(1/\tau)}$, which has a simple mode expansion:
\begin{equation}
  \MSE_{\tau,\tau}(q+\eta) = \MSE(q+\eta) - \sum_k (1 + \tau / \sigma_k) e^{-\tau / \sigma_k} \langle v_k , \eta \rangle^2.
\end{equation}
We thus see that $\MSE_{\tau,\tau}(q+\eta)$ performs well when $\tau \gg \sigma_k$, but performs poorly when $\tau \ll \sigma_k$, and therefore is insensitive to slowly decaying modes.

\section{Connection to the mean squared residual}
\label{sec:msr}

Another approach to solving for the committor is to minimize the mean squared residual (MSR) \cite{strahan_predicting_2023},
\begin{equation}
  \MSR_\tau(u) = \int (u(x) - \E[\hat{u}(X_{\tau \bmin S_0}) \mid X_0 = x])^2 \mu(x) \odif{x},
  \label{eq:msr}
\end{equation}
where $\mu$ is an arbitrary distribution of initial states.
This objective typically requires two or more independent simulations from each starting configuration $x$ to avoid bias due to the double sampling problem \cite{strahan_predicting_2023,strahan_inexact_2023}.
However, when the dynamics satisfy detailed balance, forward-in-time and backward-in-time conditional expectations are equal, and so \eqref{eq:msr} with $\mu = \pi$ can be expressed as
\begin{equation}
  \MSR_\tau(u) = \E[
    (u(X_\tau) - \hat{u}(X_{2 \tau \bmin S_\tau}))
    (u(X_\tau) - \hat{u}(X_{0 \bmax \bar{S}_\tau}))
  ],
\end{equation}
because $X_{2 \tau \bmin S_\tau}$ and $X_{0 \bmax \bar{S}_\tau}$ are independent samples conditioned on $X_\tau$.
The squared residual can be written as
\begin{equation}
  (u(X_\tau) - \hat{u}(X_{2 \tau \bmin S_\tau}))
  (u(X_\tau) - \hat{u}(X_{0 \bmax \bar{S}_\tau}))
  = g(X_{[0,\tau]},u) + g(X_{[\tau,2\tau]},u) - g(X_{[0,2\tau]},u),
\end{equation}
allowing the MSR to be expressed in terms of the EVCN loss \eqref{eq:evcn_loss}:
\begin{equation}
  \MSR_\tau(u) = 2 \tau (\tilde{L}_\tau(u) - \tilde{L}_{2\tau}(u)) = \MSE_{\tau,2\tau}(u).
\end{equation}
As discussed in Appendix~\ref{sec:Lextrapolation}, $\MSE_{\tau,2\tau}(u)$ is variational, and thus $\MSR_\tau(u)$ is variational as well.

\section{Further discussion of overfitting}
\label{sec:overfitting}

The loss functions \eqref{eq:vcn_loss} and \eqref{eq:evcn_loss} can exhibit pathological overfitting when sampling is sparse, which is common for high-dimensional systems.
When no configuration appears more than once in a dataset, highly expressive models and input features (e.g., those from Ref.~\citenum{pengmei_using_2025}) can identify the trajectory and time index of each configuration, ignoring the actual dynamics.

In the worst-case scenario, the dataset consists of a single trajectory $X_{[0,\tau]}$ from each initial structure $X_0$, with no configuration appearing more than once.
With highly expressive models and input features, the loss can be minimized independently for each trajectory.
Under these conditions, the VCN loss is minimized at
\begin{equation}
  \min_u L_\tau(u) = \frac{1}{2\tau} \E[\1_A(X_0) \1_B(X_\tau) + \1_B(X_0) \1_A(X_\tau)],
\end{equation}
with $u(X_0) = u(X_\tau)$ if $X_0 \in D$ and $X_\tau \in D$; otherwise, $u(X_0) = \1_B(X_0) + \1_D(X_0) \1_B(X_\tau)$ and $u(X_\tau) = \1_B(X_\tau) + \1_D(X_\tau) \1_B(X_0)$.
The EVCN loss is minimized at
\begin{equation}
  \min_u \tilde{L}_\tau(u) = \frac{1}{2\tau} \E[N_{AB}(X_{[0,\tau]}) + N_{BA}(X_{[0,\tau]})],
\end{equation}
with $u(X_0) = u(X_\tau)$ if $X_t \in D$ for all $t \in [0,\tau]$; otherwise, $u(X_0) = \1_B(X_{\tau \bmin S_0})$ and $u(X_\tau) = \1_B(X_{0 \bmax \bar{S}_\tau})$.
For both loss functions, when $u(X_0) = u(X_\tau)$, the output is otherwise unconstrained and can take arbitrary values without affecting the loss.

These issues can arise even with long trajectories if each configuration still appears only once.
With a single long trajectory and a single-step lag time, minimizing the loss effectively reduces to solving for the committor along a 1D random walk on the time index.
At long lag times, the EVCN loss is minimized with the empirical committor $u(X_t) = (\1_B(X_{S_t}) + \1_B(X_{\bar{S}_t}))/2$, which takes values in $\{0,0.5,1\}$.

To mitigate these problems, we treated the number of optimization steps as a hyperparameter and selected it based on validation performance.
We also restricted the model's flexibility by using a simple architecture and less expressive input features.
A more principled solution would be to explicitly regularize the model to prevent it from encoding the trajectory and time index, which we leave for future work.

\end{appendices}

\section*{Acknowledgments}

We thank D.~E.~Shaw Research for making the molecular dynamics trajectories available to us.  This work was supported by National Institutes of Health award R35 GM136381 and National Science Foundation award DMS-2054306.
This work was completed with computational resources administered by the University of Chicago Research Computing Center, including Beagle-3, a shared GPU cluster for biomolecular sciences supported by the NIH under the High-End Instrumentation (HEI) grant program award 1S10OD028655-0.

\providecommand{\latin}[1]{#1}
\makeatletter
\providecommand{\doi}
  {\begingroup\let\do\@makeother\dospecials
  \catcode`\{=1 \catcode`\}=2 \doi@aux}
\providecommand{\doi@aux}[1]{\endgroup\texttt{#1}}
\makeatother
\providecommand*\mcitethebibliography{\thebibliography}
\csname @ifundefined\endcsname{endmcitethebibliography}
  {\let\endmcitethebibliography\endthebibliography}{}

\end{document}